\documentclass[11pt]{article}
\usepackage{hyperref}
\pdfoutput=1
\begin{document}
\title{Control of a Circular Jet}
\author{Trushar B. Gohil, Arun K Saha and K. Muralidhar
\\\vspace{6pt} Department of Mechanical Engineering \\
\vspace{6pt} Indian Institute of Technology Kanpur, Kanpur 208~016, India}
\maketitle
\begin{abstract}
\noindent
The present study report direct numerical simulation (DNS) of a circular jet and the effect of
a large scale perturbation at the jet inlet. The perturbation is used to control the jet 
for increased spreading. Dual-mode perturbation is obtained by combining an axisymmetric
excitation with the helical. In the fluid dynamics videos, an active control of the 
circular jet at a Reynolds number of 2000 for various frequency ratios (both integer and 
non-integer) has been demonstrated. When the frequency ratio is fixed to 2, bifurcation of the jet on a 
plane is evident. However, for a non-integer frequency ratio, the axisymmetric jet is 
seen to bloom in all directions.

\end{abstract}
\section*{}
Fluid jets emerging from a circular tube are commonplace and more so in a variety of 
applications. The performance of a jet is to be seen in the context where it is put to
use. The jet should be highly directional if used as thruster and must spread maximally 
if used for mixing. Preliminary research shows that forcing clearly leads to large-scale
effects in the spatial development and characteristics of the flow field.
Under certain perturbation conditions, the outcome of control may result in blooming 
of the axisymmetric jet. A blooming jet shows axisymmetric spreading in all 
directions. When the frequency ratio is a non-integer,
Lee and Reynolds (1985) experimentally showed how simultaneous axial and orbital 
excitations can be used to dramatically increase the growth of the jet shear and 
formation of blooming jet. These effects are demonstrated numerically in the present study.
The simulation uses the MAC (Marker-And-Cell)
method based on higher order discretization. These are the 4$^{th}$ order central difference for
the viscous terms, a hybrid of 4$^{th}$ and 5$^{th}$ order upwind for the convective terms while the temporal 
derivative uses a 2$^{nd}$ 
order Adam-Bashforth scheme. The number of grid points used in the present simulation is
$210\times 276\times 276$. The Reynolds number based on the maximum incoming velocity and the
diameter of 
the nozzle is fixed at 2000. Dual-mode perturbation is obtained
by combining an axisymmetric excitation with the helical. 
\par
The spreading of the jet 
is found to depend solely on the ratio, $R_f$, between the non-dimensional axial 
frequency ($St_D$)and the orbital frequency ($St_H$).  The iso-surfaces of vortical 
structures ($Q$-function) have been presented in the videos for the demonstration of 
bifurcation and blooming phenomena. 
The jet bifurcates for a frequency ratio of 2 while the formation of the blooming jet 
with a varying number of branches is possible by changing the frequency ratio away from two,
as a non-integer. The details of the study are available in Gohil (2010).
 ~\\ \\
There are two videos uploaded in the arXiv system: one of low resolution (APS\_Gohil\_Saha\_Muralidhar-352x288.mpg; 
352$\times$288) and the other, with a high resolution 
(APS\_Gohil\_Saha\_Muralidhar-720x576.mpg; 720$\times$576)
 ~ \\

\noindent {\bf Reference} \\
1.	Lee M. and Reynolds W.C., Bifurcating and blooming jets, Report TF-22, 
Thermosciences Division, Department of Mechanical Engineering, Stanford University, 
Stanford, CA 1985. \\
2.    Gohil, T. B., "Control Of Circular And Square Jets Using Large Scale Perturbations: 
A Numerical Study", Ph.D Thesis, Indian Institute of Technology Kanpur, 
Kanpur, India, 2010.

%
\end{document}